\documentclass[aps,prx, 10pt,
twocolumn,
english,
superscriptaddress,
floatfix,
longbibliography]{revtex4-2}

\usepackage{physics}
\usepackage{dsfont}
\usepackage{color,newfloat}
\usepackage{graphicx}
\usepackage{dcolumn}
\usepackage{bm}
\usepackage{amsmath,amssymb,amsthm}
\usepackage{mathtools}
\usepackage{multirow}
\usepackage[dvipsnames]{xcolor}
\usepackage{enumerate}
\usepackage{booktabs}
\usepackage{colortbl}
\usepackage{xcolor}

\usepackage[colorlinks=true]{hyperref}
\hypersetup{
    colorlinks=true,
    linkcolor=Maroon,
    citecolor=Maroon,
    urlcolor=Maroon
}

\makeatletter
\def\maketitle{
\@author@finish
\title@column\titleblock@produce
\suppressfloats[t]}
\makeatother

\begin{document}

\newcommand{\TitleName}{Fusion-based implementation of qLDPC codes with quantum emitters}
\title{\TitleName}

\author{Susan X. Chen}
\email[Corresponding author: ]{susan.chen@nbi.ku.dk}
\affiliation{Quantum Engineering Centre for Doctoral Training, H. H. Wills Physics Laboratory and School of Electrical, Electronic, and Mechanical Engineering, University of Bristol, BS8 1FD, United Kingdom}
\affiliation{NNF Quantum Computing Programme, Niels Bohr Institute, University of Copenhagen, Blegdamsvej 17, DK-2100 Copenhagen Ø, Denmark}

\author{Matthias C. L\"{o}bl}
\affiliation{Center for Hybrid Quantum Networks (Hy-Q), The Niels Bohr Institute, University~of~Copenhagen,  DK-2100  Copenhagen~{\O}, Denmark}
\affiliation{Sparrow Quantum, Blegdamsvej 104A, DK-2100  Copenhagen~{\O}, Denmark}

\author{Ming Lai Chan}
\affiliation{Center for Hybrid Quantum Networks (Hy-Q), The Niels Bohr Institute, University~of~Copenhagen,  DK-2100  Copenhagen~{\O}, Denmark}
\affiliation{Sparrow Quantum, Blegdamsvej 104A, DK-2100  Copenhagen~{\O}, Denmark}

\author{Anders S. Sørensen}
\affiliation{Center for Hybrid Quantum Networks (Hy-Q), The Niels Bohr Institute, University~of~Copenhagen,  DK-2100  Copenhagen~{\O}, Denmark}

\author{Stefano Paesani}
\email{stefano.paesani@nbi.ku.dk}
\affiliation{NNF Quantum Computing Programme, Niels Bohr Institute, University of Copenhagen, Blegdamsvej 17, DK-2100 Copenhagen Ø, Denmark}

\begin{abstract}
\section*{Abstract}
Quantum low-density parity check (qLDPC) codes offer higher encoding rate than topological codes, e.g. surface codes, making them favourable for practical, fault-tolerant quantum computing with low overhead. These codes are particularly well-suited for fusion-based photonic implementations as this platform readily supports non-local connections. We propose an architecture specifically tailored to quantum emitters which can implement any Calderbank–Shor–Steane (CSS) qLDPC code. In this architecture, the photonic resource states are deterministically produced via quantum emitters and a conditional repeat-until-success strategy is incorporated to achieve high photon loss tolerance. We simulate small exemplary Bivariate Bicycle qLDPC codes and analyse the performance of our constructions under relevant physical noise mechanisms, including erasures due to fusion failure or photon loss, as well as Pauli errors. We obtain performances comparable with topological architectures though with significantly higher encoding rates.
\end{abstract}

\date{\today}

\maketitle

\section{Introduction}

The development of reliable quantum computers critically depends on advancing fault-tolerant quantum computing architectures that provides access to large numbers of error-corrected qubits with practical quantum hardware. For the past decades, topological codes, such as the surface code, have been studied extensively from a theoretical perspective and have become the standard platform for developing these architectures. This is due to their local connections in a planar layout and their high error-correction thresholds~\cite{Bravyi1998, Dennis2002, Breuckmann2021}. These codes have also recently been experimentally demonstrated in fault-tolerant regimes~\cite{Google2024}. Despite this, they suffer from very limited encoding rate, where each surface code patch is only able to encode a single logical qubit and the associated overhead to access large numbers of error corrected logical qubits pose a notable challenge for current and near-term quantum hardware. 

In recent years, general (non-topological) classes of low-density parity check (qLDPC) codes with significantly improved encoding rates and similar error correction performances have been developed~\cite{Breuckmann2021, bicyclecodes}. However, this improvement comes with the challenge that very often long-distance connectivity is required when the qubits are arranged on a planar two-dimensional geometry~\cite{bicyclecodes, Breuckmann_2016}, posing difficulties for implementations in static solid-state qubit platforms, such as superconducting qubits and gate-defined quantum dots. Meanwhile, such long-range interactions can be implemented with other platforms. For example, photonic qubits can be transmitted over long distances using waveguides and optical fibres, and atomic qubit arrays are reconfigurable to enable interactions between distant sites~\cite{alexander2024manufacturableplatformphotonicquantum, Bluvstein_2023}. 

With such capabilities, these platforms can naturally benefit from the more efficient encoding of qLDPC codes, and architectures for atomic qubits have already been explored based on two-qubit gates between Rydberg atomic states~\cite{xu2023}. Photonic qubits, however, lack direct qubit-qubit interactions and thus require significantly different constructions. One possibility encodes qubits into bosonic modes, enabling deterministic entangling interactions. Such an architecture with arbitrary qLDPC code input was recently proposed in Ref~\cite{PhysRevLett.134.100602}. However, it relies on GKP states which are experimentally challenging to generate and on operations tailored specifically to continuous-variable systems. An alternative approach uses single-qubit encodings, along with probabilistic two-qubit operations, so-called \textit{fusions}~\cite{loqc, thomas2024fusion}. 

Fusion-based quantum computing (FBQC) has recently been developed as a method to build and analyse fault-tolerant constructions based on these gates, where operations are performed by consuming small entangled photonic resource states through fusions~\cite{fbqc, PhysRevA.70.060302}. In this article we construct fusion-based logical memories based on qLDPC codes that offer better encoding rates, specifically Bivariate Bicycle codes~\cite{bicyclecodes}, and analyse their thresholds with phenomenological noise models as well as photon loss.  This approach generalises to any Calderbank–Shor–Steane (CSS) qLDPC code and implements a logical memory solely through photonic fusions and resource states that can be deterministically generated via quantum emitters~\cite{PhysRevA.82.032332,Paesani2023}.

\begin{figure*}[ht]
    \centering
    \includegraphics[width=0.99\textwidth]{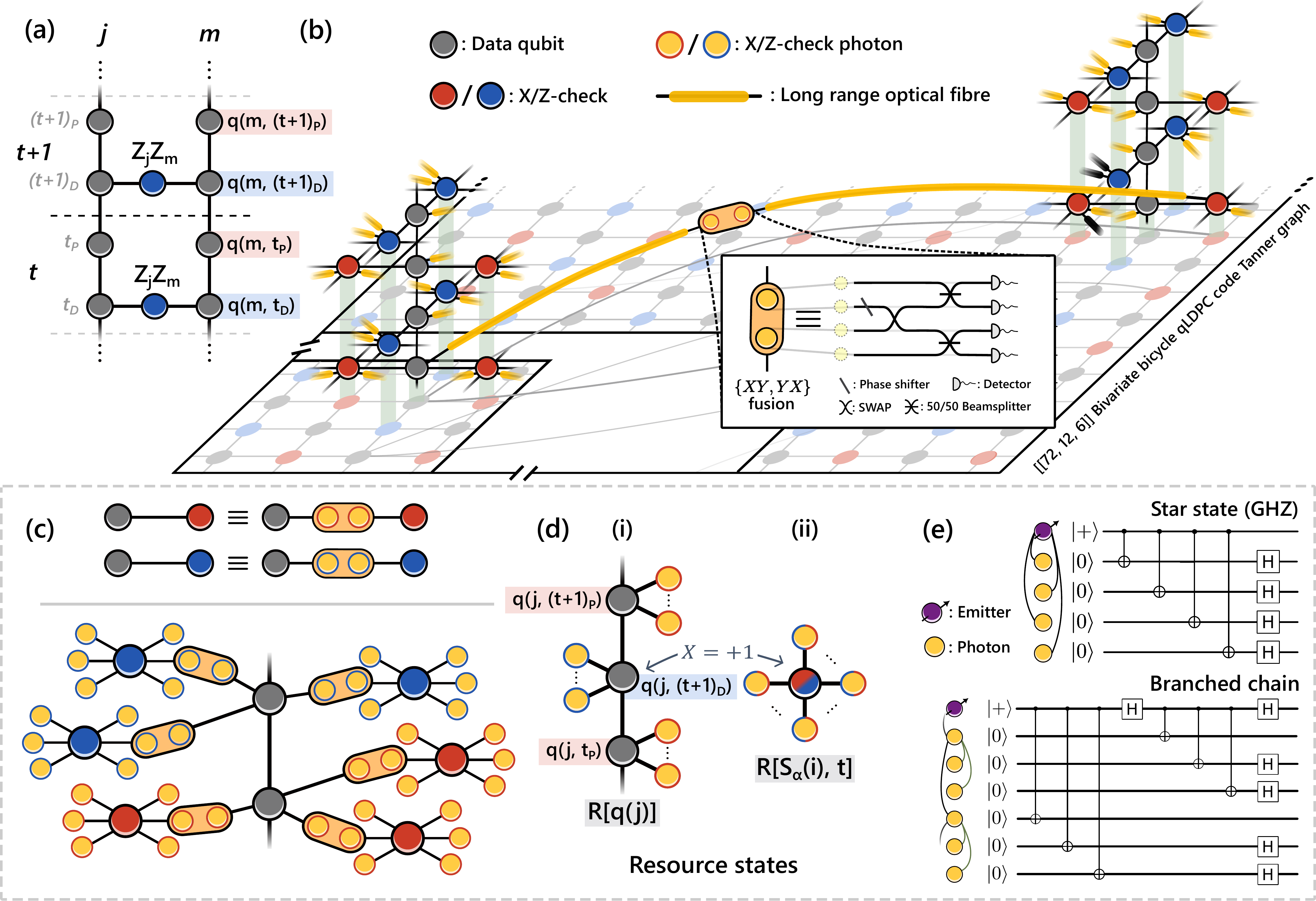}
    \caption{\textbf{Outline of the architecture.} (a) Setup for the measurement of parity check $Z_jZ_m$ of two foliated data qubits $j$ and $m$ at time intervals $t$ and $t+1$ by coupling their $t'_D$ layer qubits to ancillary qubits (blue). In between the $Z$-parity checks, parity check of the $X$-operators are performed with other data qubits. (b) Foliated cluster state lattice of the $[[72, 12, 6]]$ Bivariate Bicycle qLDPC code (partially shown in a Toric layout) from Ref.~\cite{bicyclecodes}, where each \textit{layer} constitutes only $X$ or $Z$-checks upon measuring the qubits. In addition to the four nearest neighbour links, each qubit also has two long-range connections, such as the ones shown for a sub-set of nodes in the Tanner graph. In our fusion-based decomposition of the lattice, each data qubit to ancilla qubit link is split and realised with an $\langle XY, YX\rangle$ photonic fusion for which a setup is shown in dual-rail encoding. (c) An illustration of how a data qubit completes its one error detecting round (layer) of checks in the lattice via fusions. (d) The two types of resource states that make up the fusion lattice. (i) Data branched chains:  Linear cluster states of length $2T$ with $\eta_Q/2$ leaves per layer. (ii) Ancilla GHZ states: Shown as $(\eta_A + 1)$-star graph states for clarity. We note that all qubits which are not explicitly drawn as fusion photons (yellow) are projected onto the measurement outcome $X=+1$. (e) Deterministic generation protocols for the star and branched chain photonic graph states using quantum emitters as detailed in Refs.~\cite{lindner-rudolph, Paesani2023}.}
    \label{fig:diagrams}
\end{figure*}

\section{Results}
\subsection{qLDPC Codes}

Quantum low-density parity check codes are stabiliser codes with low and constant qubit and check degrees. As a result, these codes are described by sparse parity check matrices~\cite{Breuckmann2021}. Following standard notation, we present codes in terms of their parameters $[[n,k,d]]$, where $n$ is the number of physical qubits, $k$ is the number of encoded logical qubits, and $d$ is its distance, i.e. the smallest possible Hamming weight of a logical operator. We focus on qLDPC codes that are CSS where checks are made up of solely Pauli $X$- or $Z$-operators and therefore allow the two types of error to be detected and corrected independently.

Recently, qLDPC code families with encoding rates $k/n$ greater than those of topological codes have been studied~\cite{hyperbicycle_codes, hypergraph_product_codes}. It has also been shown that there exist codes with \textit{asymptotically good} parameters~\cite{Panteleev2022, quantum_tanner_codes}, i.e. codes for which the number of encoded qubits and the distance both scale linearly with the number of physical qubits. Such codes offer a clear route to reducing overhead in fault-tolerant schemes.

We consider the recently proposed Bivariate Bicycle qLDPC code family~\cite{bicyclecodes} (See Supplementary Information III). These are the first known examples of codes that can compete with the surface code in terms of logical error rate while offering a higher encoding rate. The choice to focus on Bivariate Bicycle codes is arbitrary, and serve solely as an example study. Our approach is applicable to any qLDPC CSS code and could therefore also be applied to a larger class of codes with proven asymptotically constant encoding rate~\cite{quantum_tanner_codes, Panteleev2022}.

\subsection{Construction}
Measurement-based quantum computing models temporally evolve the encoded logical information by performing single-qubit measurements on a large entangled graph state. A graph state, $\ket{G}$, is represented by an undirected graph $G=(V,E)$ where the vertices $V$ are qubits initialised in the state $\ket{+}=\frac{1}{\sqrt{2}}(\ket{0}+\ket{1})$ and edges $E$ are controlled-phase operations $U_c^{a,b}$ between the vertices $a,b$ that they connect:

\begin{equation}
    \ket{G}=\prod_{a,b\in E} U_c^{a,b} \ket{+}^{\otimes V}
\end{equation}
Graph states are $+1$ eigenstates of their stabiliser generators 
\begin{equation}
    K_i = X_i\prod_{j\in N_i}Z_j ,
\end{equation}
where $N_i$ is the neighbourhood of qubit $i$~\cite{Hein2004}, and $X_i$ and $Z_j$ are Pauli operators.

Typically, fault-tolerant circuit-based implementations of a CSS \footnote{Foliation can also be applied to non-CSS codes~\cite{foliation}, though in the rest of this manuscript we will assume CSS constructions.} code require $d$ repetitions of its stabiliser measurements. This temporal sequence of operations can be replicated in a measurement-based sense by \textit{foliating} the underlying \textit{base} error-correcting code~\cite{Bolt2016, cjb4-l57n}. Foliation transforms the code into a graph state with a layered structure, replacing time with an extra spatial dimension perpendicular to the layers. (We will also refer to this graph state as a cluster state or lattice in the following). Using the cluster state, one can perform all repeated check measurements by measuring the state layer by layer~\cite{foliation}. As illustrated in Fig.~\ref{fig:diagrams}(a), this graph state has alternating layers of bipartite graphs where data qubits would be connected to ancilla qubits according to the structure of the Tanner graph associated with the $Z$- and $X$-type parity checks of the underlying CSS code. The data qubits are connected between these layers~\cite{Bolt2016}. This structure encodes data qubits of the base code in a one-dimensional chain of length $2T$ qubits, where we label the $j^{th}$ such data qubit as $q(j)$ and $T$ is the number of stabiliser measurement rounds of the logical memory. The logical information of the chain is locally accessible and can be propagated along the chain, analogously to temporal evolution, by performing single-qubit $X$-measurements on \textit{past qubits} of the chain.  
Figure~\ref{fig:diagrams}(a) shows an example of the repeated measurement of a small parity check between two foliated data qubits, $j$ and $m$.
At each time-step layer, stabilisers are measured by measuring their ancilla qubits in the $X$-basis. Due to the way chain logical operators are accessed, the $Z$ and $X$ stabilisers are measured sequentially, so two "temporal" layers  $t=\{t_D,t_P\}$ are required to complete one full measurement round of the code. Throughout the rest of the paper, we will refer to the layer in which only $Z$-stabilisers are measured as the dual layer, $t_D$, and the layer in which $X$-stabilisers are measured as the primal layer, $t_P$. The physical qubits on the chain from the $j^{th}$ base data qubit are labelled $q(j,t_D)$ and $q(j,t_P)$.

The cluster state resulting from the procedure described above possesses graph state stabilisers by definition. By multiplying graph state stabiliser generators, one finds that there exists two sets of regularly structured local detectors spanning three layers of the foliated lattice~\cite{Bolt2016}. These detectors form \textit{unit cells} of two separate lattices: $X$-type and $Z$-type (generalisations of primal and dual lattices in topological code foliations). 

The $X$-type detectors detect changes in parity of the $X$-checks of the base error-correcting code labelled $S_X(i)$, and thus errors which may contribute to a logical $X$ error in the foliated code. We denote $\mathcal{X}_{i,t}$ as the set of qubits on which there is Pauli-X operator support for the detector associated with the $i^{th} X$-check at time interval $t$:

\begin{align}
\mathcal{X}_{i,t} :=\ &\underbrace{\{ s_X(i,t_P) \}}_{\text{1st layer}} 
\cup \underbrace{\{ q(j,(t+1)_D) \mid j \in S_X(i) \}}_{\text{2nd layer}} \notag \\
&\cup \underbrace{\{ s_X(i,(t+1)_P) \}}_{\text{3rd layer}}.
\end{align}
Here, we label $ s_X(i,t_P)$ as the physical ancilla qubit associated with the $i^{th} X$-check $S_X(i)$ of the base code within layer $t_P$ of the foliated lattice, and $q(j,(t+1)_D)$ as the physical data qubit in the dual layer at the time interval $t+1$ corresponding to the $j^{th}$ foliated chain. A logical $X$ operator of the cluster state, or \textit{correlation surface}, is effectively a surface that traverses the foliated length $2T$ of the lattice. This surface has weight on chain qubits associated with the corresponding, say $k^{th}$, logical $X$ operator of the base code $\bar{X}_k^{base}$ across all $t_D$ layers. In other words, the $k^{th}$ logical $X$-operator $\bar{X}_k$ for the lattice is

\begin{equation}
    \bar{X}_k=\prod_{t_D=1}^T \bar{X}_k^{base}(t_D).
\end{equation}

Analogously, a $Z$-type detector cell helps detect physical errors which may contribute to logical $Z$ errors and has a comparable structure, spanning the layers $\{t_D, t_P, (t+1)_D\}$:

\begin{align}
\mathcal{Z}_{i,t} :=\ &\underbrace{\{ s_Z(i,t_D) \}}_{\text{1st layer}} 
\cup \underbrace{\{ q(j,t_P) \mid j \in S_Z(i) \}}_{\text{2nd layer}} \notag \\
&\cup \underbrace{\{ s_Z(i,(t+1)_D) \}}_{\text{3rd layer}}.
\end{align}
The $k^{th}$ logical $Z$-operator $\bar{Z}_k$ is similarly defined 

\begin{equation}
    \bar{Z}_k=\prod_{t_P=1}^T \bar{Z}_k^{base}(t_P).
\end{equation}

Using the method in Ref.~\cite{Bolt2016}, we may foliate any CSS qLDPC code with parity check matrices $H_X$ and $H_Z$. Here, we consider codes that, in addition, have constant qubit-degree $\eta_Q$ and constant check-degree $\eta_A$. This is not required for our approach to apply but simplifies the fusion-based construction \footnote{Otherwise, variations of the two type of resource states (see below) are required.}. The qubit-degree is the number of checks each data qubit is connected to, and check-degree is the number of data qubits involved in each check. For Bivariate Bicycle codes, $\eta_Q =\eta_A=6$ with every qubit being involved in three $X$ and three $Z-$type checks. In our analysis of the codes, we choose the number of measurement rounds,  $T$, to be the error distance $d$ of the underlying code.

To reduce the demands of producing the lattice as a large entangled cluster state, we consider a fusion-based approach where the foliated system is decomposed into smaller resource states and type-II fusion operations~\cite{loqc} between them (shown in Figure~ \ref{fig:diagrams}(b) for the $[[72,12,6]]$ Bivariate Bicycle qLDPC code). Fusions are probabilistic Bell state measurements on two qubits $A, B$ that measure joint parities such as $\langle X_AX_B, Z_AZ_B \rangle$ upon success~\cite{Gimeno2016, Lobl2024b, thomas2024fusion} (other parity measurements can be realised via rotation with single-qubit gates~\cite{Gimeno2016} $-$ see Supplementary Information II for details). Naively, these operations succeed with a $50\%$ probability in photonics~\cite{loqc}, however, this probability can be boosted with the addition of ancillary photon pairs as part of the fusion gate~\cite{Grice2011, Ewert2014, hauser2025boosted}. 

In the considered foliated structure, the data qubits are connected between different layers, forming chain structures. Within every layer, the graph is bipartite as data qubits are only connected to ancilla qubits (see Fig.~\ref{fig:diagrams}(c)). This structure gives rise to a very natural fusion-based construction using branched chain graph states (caterpillar trees in graph theory~\cite{Harary1973}) which are linear cluster states with additional \textit{leaf} qubits connected to qubits of the linear chain, and star-shaped graph states. The data qubits form the central path of branched chains (grey in Fig.~\ref{fig:diagrams}(c,d)), and ancilla qubits form the central qubits of star-shaped graph states (red, blue in Fig.~\ref{fig:diagrams}(c,d)) where they are connected to the branched data qubit chains by fusions (yellow in Fig.~\ref{fig:diagrams}(c,d)). Importantly, branched chains constitute exactly the class of states that the scheme from Ref.~\cite{lindner-rudolph} can generate using a single quantum emitter~\footnote{Furthermore, by applying Hadamard gates on the leave qubits of a branched chain it is converted into a redundantly encoded linear cluster state~\cite{PhysRevA.82.032332,Paesani2023} which enables using encoded rather than physical fusions~\cite{chan2024tailoring}.}. This approach employs a sequence of spin gates on a quantum emitter and optical excitation which induces photon-emission from the emitter~\cite{PhysRevA.58.R2627}. Figure~\ref{fig:diagrams}(e) shows the quantum circuit that deterministically generates a branched chain and a star-shaped graph state, which is a branched chain of length one and is locally equivalent to the GHZ. Below we give a more detailed description of the involved resource states 1) data branched chains, and 2) ancilla GHZ states.

\begin{figure*}[ht]
    \centering
    \includegraphics[width=\textwidth]{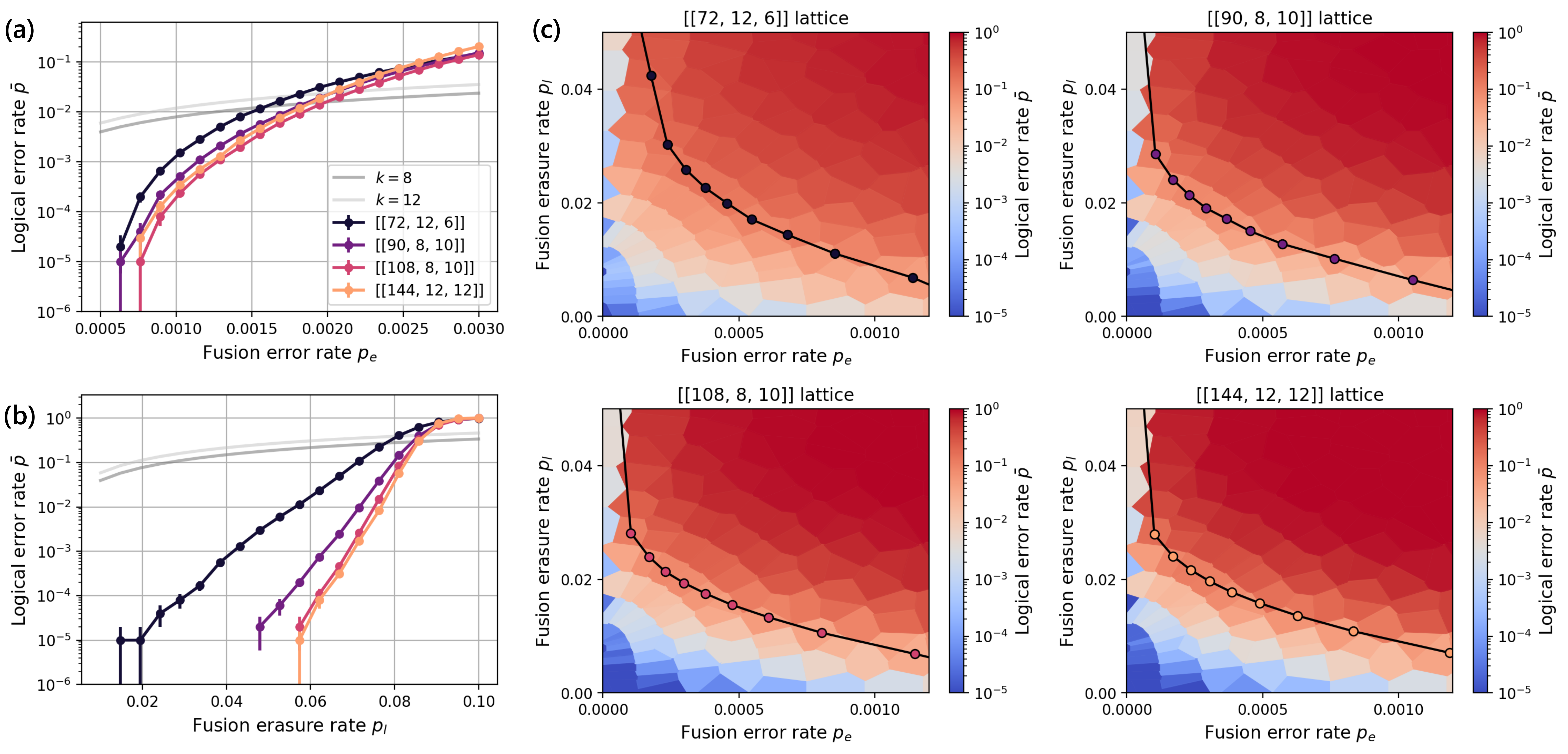}
    \caption{Logical error rates $\bar{p}$ for lattices constructed from small examples of Bivariate Bicycle qLDPC codes given (a) fusion error only and (b) fusion erasure only. The grey curves labelled $k=8$ and $k=12$ represent the probability of having at least one error across the $k$ physical qubits at a given physical noise rate, i.e., $p_k^{phys}=1-(1-p)^k$. The pseudo-threshold is defined as the point where the logical error rate of the lattice intersects with its corresponding grey curve, a.k.a., $\bar{p}=p_k^{phys}$. Lower bounds on the truncated error bars in Figure a) and b), not shown for ease of visualisation, are $5 \cdot 10 ^{-11}$. (c) Logical error rate as a function of fusion erasure $p_l$ and error $p_e$ rates  for each lattice. The points on each colour plot correspond to the said lattice's pseudo-threshold surface, such that the area below the curve can be interpreted as a \textit{pseudo-correctable region}. }
    \label{fig:logical error rates}
\end{figure*}

\textit{Data branched chains} are branched chains of length $2T$ with $\eta_Q/2$ branches stemming from each physical qubit of the chain. Being our equivalent of a \textit{foliated qubit}, the lattice contains one data branched chain per underlying code qubit, hence we label these resource states $R[q(j)]$ where $q(j)$ is the $j^{th}$ base code data qubit. The branched qubits of $R[q(j)]$ at layer $t_D$($t_P$) will go into fusions which complete all the $Z$($X$) code checks that the code data qubit $q(j)$ takes part in. All grey qubits in Fig.~\ref{fig:diagrams}(d)(i) are virtual qubits assumed to be measured in the $X$-basis with the measurement outcome $+1$ as discussed below.

\textit{Ancilla GHZ states} are visually represented, in Figure~\ref{fig:diagrams}(d)(ii), as star graphs with $\eta_A$ leaves with the central qubit measured and projected into $+1$ in the $X$-basis. It can be shown that the resulting state from this requirement is a GHZ state on  $\eta_A$ qubits. The lattice comprises one ancilla-GHZ state per code check involved in each layer which we label $R[S_\alpha(i),t]$, where $S_\alpha(i)$ denotes the $i^{th}$ base code check of type $\alpha$ $\in \{X,Z\}$ as above and $t$ is the time interval at which the check takes place. These ancilla states fuse with the leaf qubits of the data resource states that are part of the check (Figure~\ref{fig:diagrams}(c) $-$ for details of choice of fusion type see Supplementary Information II).

We remark that the central qubits of the branched chain and $(\eta_A + 1)$-star graph are virtual~\cite{fbqc, Bell2023optimizing}: they are used to simplify the visualisation of the graph, but in practice the physical state to be generated corresponds to the graph obtained measuring the virtual qubits in $X$ and with outcome +1 (see Supplementary Information I for a depiction of the reduced resource states and resulting lattice). We  note that the state after applying a Pauli measurement on a branched chain is local Clifford equivalent to another branched chain \footnote{This is because a Pauli measurement creates a vertex minor~\cite{Dahlberg2018} of the graph and a vertex minor cannot have an increased linear rank width.}. Using the scheme from Ref.~\cite{lindner-rudolph}, one thus can directly generate the state resulting from measuring the central qubits of the branched chain or $(\eta_A + 1)$-star graph in the $X$-basis. In other words, these virtual qubits do not need to be generated in the first place and fusions are the only measurements that need to be applied, giving rise to an entirely fusion-based scheme. Since photons that are not generated cannot be lost, this scheme will have a higher erasure threshold. 

When quantum emitters are employed,  it turns out that the above fusion-based scheme does not represent the optimum strategy to reduce the effect of photon loss. More practical is likely to be a scheme where not all fusion photons are generated at once but rather on demand and fusions are repeated until success (see Supplementary information I).

\subsection{Threshold Analysis}

Using the method detailed above, we construct lattices of varying sizes of Bivariate Bicycle qLDPC codes. We test the lattices' tolerance to both error and erasure of fusion outcomes. An independent and identically distributed (i.i.d.) fusion noise model is used~\cite{fbqc}, where logical error probabilities are determined based on a physical noise rate $p$ occurring independently for each fusion based on an error probability $p_e$ and an erasure probability $p_l$. For decoding, we use a modified version of the union-find decoder from Ref.~\cite{uf-decoder} since it works for non-topological qLDPC codes and can handle Pauli errors and erasures simultaneously (see section~\ref{suppl_decode} for more details).

Topological qLDPC codes, like the surface code, have a well-defined error threshold: for error rates below the error threshold, an arbitrary low logical error rate can be obtained by increasing the code size. For general qLDPC codes, however, there is no systematic way of scaling the size of the code. Therefore, we evaluate the fault-tolerant performance of the qLDPC lattices with respect to their so-called \textit{pseudo-thresholds}. A pseudo-threshold represents the physical error rate at which the probability for an error on one out of $k$ logical qubits (the logical error rate $\bar{p}$) equals $p_k^{phys}=1-(1-p)^k$, the probability that at least one out of $k$ physical qubits has an error – this we refer to as the \textit{break-even equation}. Below the pseudo-threshold the logical error is smaller than the physical, $p_k^{phys}>\bar{p}$, making it beneficial to use encoded rather than physical qubits. In general, the noise is a combination of fusion error and erasure. We consider a break-even equation where $p = (1-p_l)p_e + p_l/2$ (see section~\ref{suppl_decode} for details). This represents the probability $p$ of an error occurring, either without erasure or due to an erasure, where erasures introduce errors with a probability of $50\%$~\cite{Delfosse2021}.  

Pseudo-thresholds are found by performing Monte Carlo simulations. We present the logical error rates for Bivariate Bicycle code lattices in Figure~\ref{fig:logical error rates}(a-b), under independent sweeps of phenomenological fusion error and erasure probabilities, and plot the break-even equations for different number of encoded qubits $k=8,12$ as grey curves. Each logical error rate point is the mean over $10^5$ independent simulation samples, and the error bars shown represent $\pm$ 1 standard deviation considering Poissonian noise due to the finite number of samples. The colour plots in Figure~\ref{fig:logical error rates}(c) show logical error rates for each lattice sweeping through error and erasure noise combinations in the grid (while also including the additional data from pseudo-threshold analysis below) where the points mark a pseudo-threshold curve for the corresponding lattice. In order to obtain these points, we introduce a parameter $x$, which we call the \textit{noise scaling factor},  that linearly parameterises a line at equally distributed set of angles from the origin. At each angle, we recover logical error rates and pseudo-thresholds by sweeping over $x$ and calculating the intersection with the modified break-even equation discussed above. As expected, it can be seen from Figure~\ref{fig:logical error rates}(a-b) that as the code size increases, the pseudo-thresholds for fusion error and fusion erasure both increase, with the $[[144, 12, 12]]$ code having an error pseudo-threshold of $0.2\%$ and an erasure pseudo-threshold of $9\%$. We note that pseudo-thresholds should not be taken as a definitive performance metric. In particular, while the $[[72, 12, 6]]$ qubit code has an unusually high comparative pseudo-threshold, its logical error suppression at low physical error rates is worse despite this.  

\subsection{Repeat-until-Success (RUS) Fusions}
\begin{figure}[ht]
    \centering
    \includegraphics[width=0.48\textwidth]{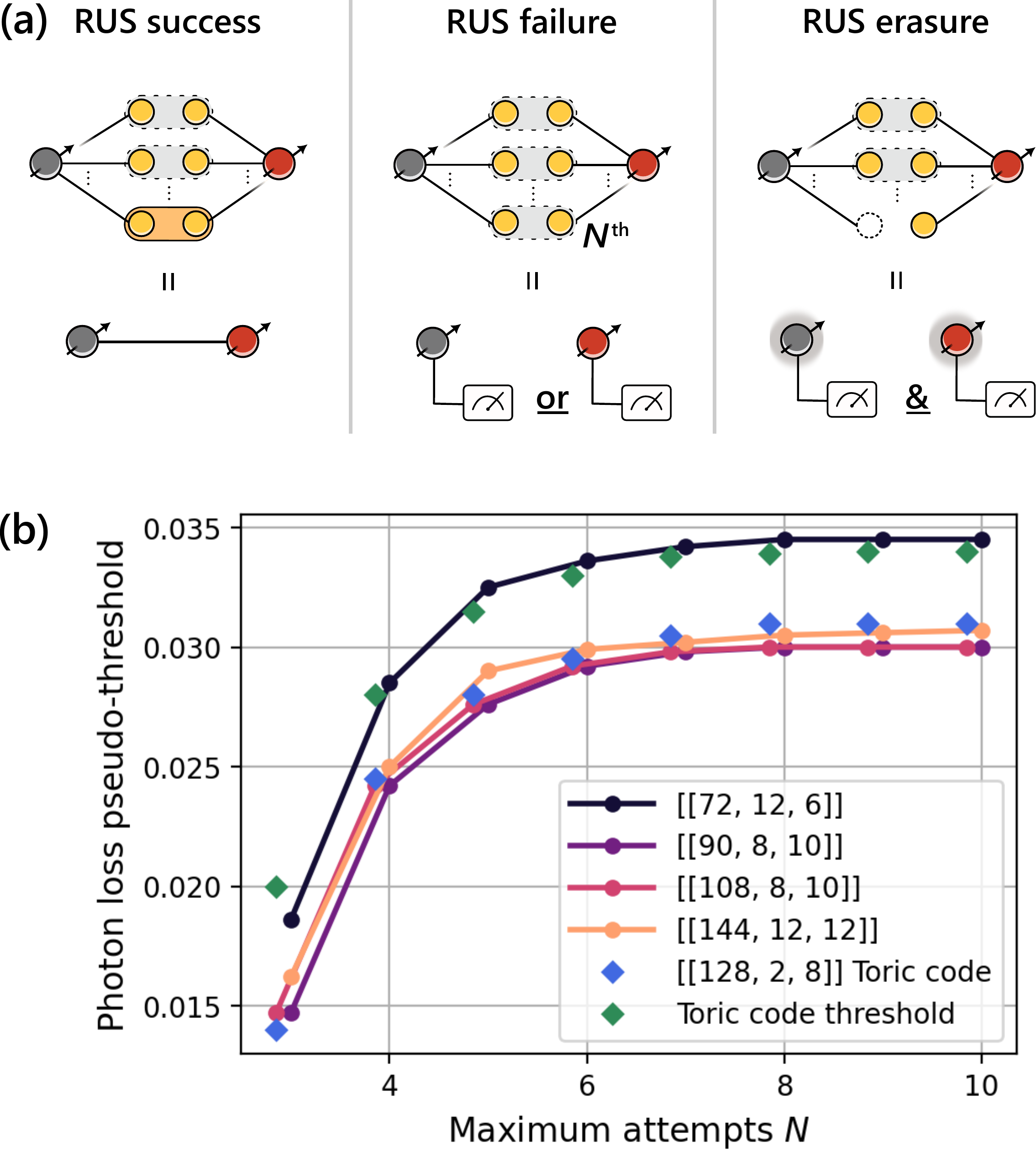}
    \caption{(a) Modified repeat-until-success strategy where RUS success, failure, and erasure result in no action, measurement of either spin, or measurement of both spins, respectively. Whether neighbouring RUS fusions are performed is adaptively decided based on the RUS fusion outcome (see Supplementary Information I). (b) Photon loss pseudo-threshold as a function of the maximum number $N$ of repetitions in a RUS fusion, shown for small Bivariate Bicycle qLDPC code lattices. The blue points show the pseudo-threshold for $[[128, 2, 8]]$, a comparable instance of the Toric code lattice, and the green points show the actual threshold for Toric code lattices. }
    \label{fig:RUS}
\end{figure}

Above, we have used abstract erasure $p_l$ and error $p_e$ rates, referring to the results of the fusions. A different question is how these quantities relate to the underlying physical parameters. To investigate this, we evaluate the tolerance of physical lattice constructions to photon loss, as it is the dominant source of noise in photonic platforms. The phenomenological erasure thresholds obtained above are too low for an implementation with standard linear optics Bell-state measurements~\cite{loqc} that fail with 50\% probability. To overcome this issue, it is thus necessary to boost the fusion probability. To achieve boosted fusion, a particularly promising technique for quantum emitters is a repeat-until-success (RUS) scheme, repeating physical fusions until success~\cite{Gliniasty2024, Lobl2024, chan2024tailoring}. At each layer, the quantum emitters stay part of the resource states and emit photons that undergo fusion measurements until the fusion is successful, lost, or a maximum number of attempts $N$ is reached and all of the physical fusions have resulted in failure (see Figure~\ref{fig:RUS}(a)). We note that fusions only occur within independent layers of the foliated structure. This enables applying the emitter-based repeat-until-success fusions with the capability for synchronous scheduling~\cite{chan2025practicalblueprintlowdepthphotonic}. RUS fusion outcomes can be sampled based on the standard analytical model described in Ref.~\cite{Gliniasty2024} where all fusions within a given layer are done in parallel, which we refer to as \textit{standard RUS}. We consider a \textit{modification} of standard RUS such that the fusions are instead performed sequentially. Note that also strategies in between are conceivable, where RUS fusions are grouped and RUS fusions of the same group are executed in parallel while the groups are executed sequentially.

The two above strategies can be understood in the picture of cluster states and the effects of RUS erasures and failures on bonds in the built cluster state lattice. \textit{RUS erasure} causes two detector cells in both the $X$-type and $Z$-type syndrome lattices to merge into a higher-weight \textit{supercell}~\cite{PhysRevLett.105.200502}. Concretely, the detector cells are merged that share the bond that the successful RUS fusion would generate between two spins. In practice, both spins are measured in the $Z$-basis (see Figure~\ref{fig:RUS}(a) RUS erasure). This is necessary because is unclear if the fusion photon connected to the left or right spin was lost, and losing one fusion photon has thus the same effect as losing both fusion photons. \textit{RUS failures} equate to missing bonds and are less detrimental because only a merge in one of the two lattices is required, and only one of the spins associated with the failed fusion must be measured in the $Z$-basis~\cite{PhysRevA.97.030301} (see Figure~\ref{fig:RUS}(a) RUS failure). While standard RUS will always measure out the spin associated with the ancilla qubit, in the modified strategy, we randomise the choice of measuring either end of the fusion to help balance cell merges across both syndrome lattices~\cite{PhysRevA.97.030301}. In addition, by performing fusions sequentially instead of simultaneously, we use the outcomes to decide about the necessity of subsequent neighbouring fusions. To increase loss tolerance, unnecessary future fusions are omitted, in particular, a fusion may be skipped if either of the spins involved needs to be measured in the $Z$-basis because of a missing bond (due to a previously failed RUS fusion). We note that a more detailed explanation of both the standard and modified variants of RUS can be found in Supplementary Information I.

With fusion outcomes sampled based on our modified RUS strategy, we perform Monte Carlo simulations on $10^5$ trials for fusion lattices built from various sizes of Toric and Bivariate Bicycle qLDPC codes. We present the photon loss pseudo-thresholds for Bivariate Bicycle code lattices in Figure~\ref{fig:RUS}(b). Similar to the previous section, we compute the pseudo-thresholds as the photon loss probabilities, which incur errors of at least one of the $k$ logical qubits with the same probability as for the physical qubits. For the $[[144, 12,12]]$ code lattice, the pseudo-threshold saturates around 3$\%$ at $N = 8$. Such photon loss values may be reached with future developments of photonic platforms~\cite{alexander2024manufacturableplatformphotonicquantum}. In Figure~\ref{fig:RUS}(b), we compare the results to the pseudo-threshold for photon loss of the $[[128, 2, 8]]$ instance of the Toric code. This is the Toric code instance closest in qubit count to the other codes, enabling a direct comparison. The photon loss pseudo-threshold for $[[128, 2, 8]]$ is seen to almost entirely overlap with those of the qLDPC codes (with the exception of the $[[72, 12, 6]]$ code). Considering that the latter allow for much larger numbers of encoded qubits, this makes them highly attractive provided that experiments can be advanced to reach the thresholds. 

We also show the actual threshold for Toric code lattices (in green), not to be confused with pseudo-thresholds at fixed code sizes, where it levels off to a maximum at 3.4$\%$ for $N=8$.  Due to our modified RUS procedure this is higher than the RUS loss threshold quoted in Ref.~\cite{Gliniasty2024} of $2.75\%$ for an equivalent implementation of the Toric code.

\section{Discussion}
We have presented an explicit method to construct cluster state lattices for fusion-based quantum computing (FBQC) from arbitrary CSS qLDPC codes. For a given input error-correcting code, the resulting lattice implements a logical memory by performing projective two-qubit fusion measurements between two distinct types of resource states. Although FBQC schemes are platform-agnostic~\cite{fbqc}, we focus on photonics where fusion operations between qubits originating from different parts of the lattice can easily be implemented by routing photons to detectors through optical fibres. The two families of photonic resource states required in this particular decomposition are readily accessible as they have been experimentally generated through deterministic spin-optical excitation protocols~\cite{Thomas2022,Cogan2023,Meng2024,Huet2025}. 

Furthermore, we employ a conditioned repeat-until-success strategy to boost photon-loss thresholds. This upgrades physical fusions to encoded fusions and uses local information from previous RUS outcomes to inform whether to perform subsequent surrounding fusions. This adaptive strategy avoids unnecessary fusions and thus reduces photon loss. Note that other threshold boosting strategies, such as dynamic biasing of failure bases based on global information have also been proposed in the literature~\cite{bombín2023increasingerrortolerancequantum}. Combining these strategies with our approach is an interesting next step to improving the loss tolerance of such constructions.

In general, qLDPC codes are highly attractive as they allow much more efficient encoding of information with dramatically reduced overhead in resources. The fact that this only comes with a modest increase in (pseudo-)threshold compared to the Toric code is a major asset of the considered approach. So far we have only considered a small subset of qLDPC codes. Since qLDPC codes are much less studied than, e.g. topological codes, it is likely that one can find codes that are more efficient than the ones considered here. It would be highly interesting to explore the full space of other qLDPC codes which offer better encoding rates and distances. Potentially, this could bring the threshold even closer to the Toric codes or reduce the resource overhead. Finally, with a particular physical platform in mind, incorporating more realistic noise models, e.g. spin errors, into simulations such as those in Ref.~\cite{chan2025practicalblueprintlowdepthphotonic} would provide important guidelines for future hardware developments towards the goal of realising fault-tolerant photonic quantum computers.  

\section{Methods}

\subsection{Details of numerical simulations} \label{A: numerics}

The pseudo-threshold points (Figure~\ref{fig:logical error rates}) were found by simulating fusion lattices of size  $T=d$ with periodic boundary conditions under different noise settings. We consider only the $X$-type lattice in these simulations due to the $Z$-type being equivalent (the parity check matrix $H_{X/Z}$ equivalence for all Bivariate Bicycle qLDPC codes implies that the distance for $X$ and $Z$ errors is the same~\cite{bicyclecodes}). 
 
\begin{figure*}[ht]
    \centering
    \includegraphics[width=\textwidth]{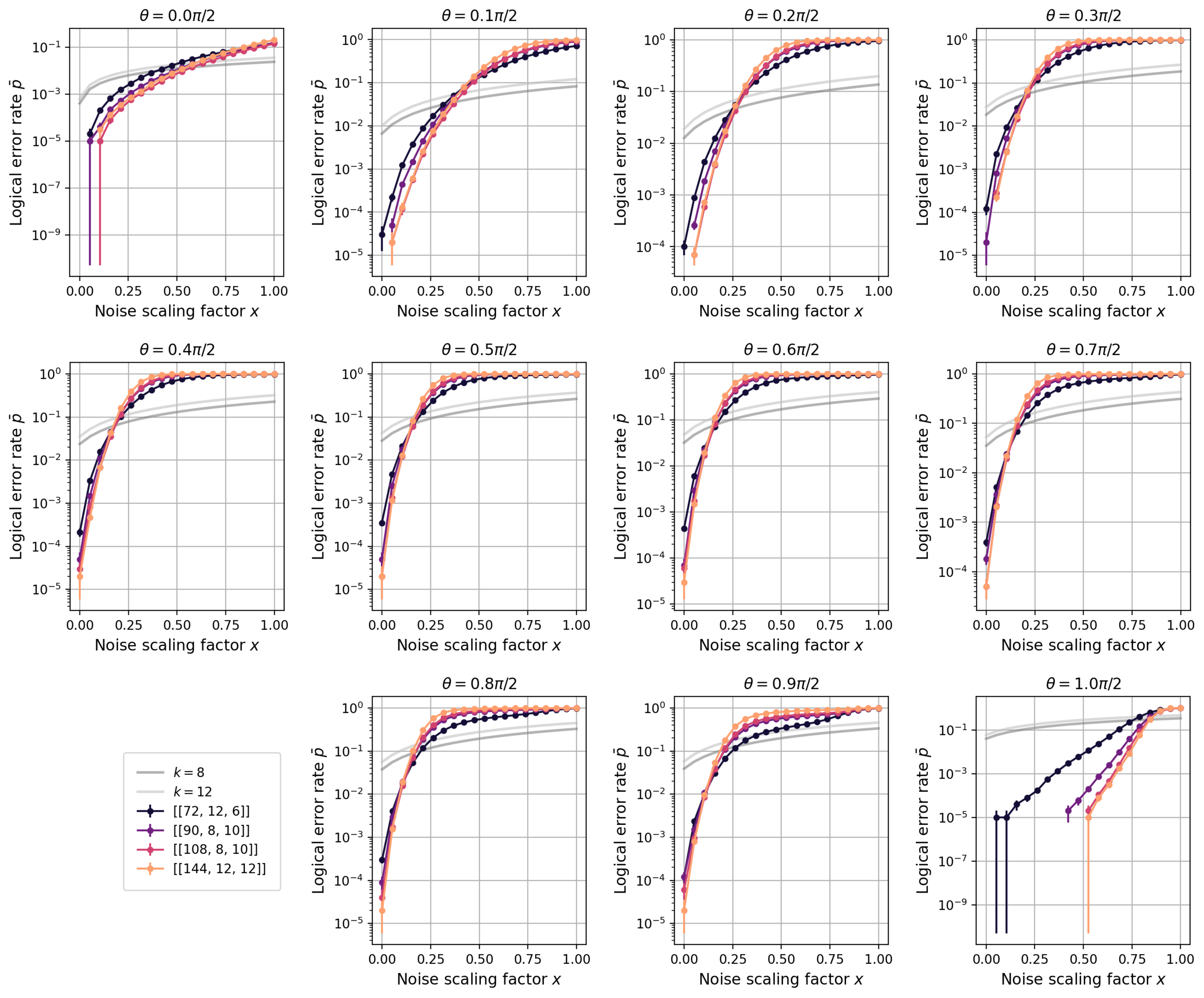}
    \caption{Logical error rates $\bar{p}$ against linearised noise parameter $x$ for lattices constructed from small examples of Bivariate Bicycle qLDPC codes, with $(p_e, p_l)\propto(x\tilde{p_e}\cos{\theta}, x\tilde{p_l}\sin{\theta})$ up to additive offsets determined by minimum probabilities $p_e^{min}$ and $p_l^{min}$. The grey curves are the break-even equations for $k=8,12$ where its intersection with the data is the pseudo-threshold. The figure shows the series of noise combination sweeps of the error/erasure phase space given by angle $\theta$, where $\theta=0$ is the error-only case and $\theta=\pi/2$ is the erasure-only case.}
    \label{fig:all_angles_ler}
\end{figure*}

We analyse performances by varying the fusion error $p_e$ and erasure rates $p_l$, which together define a two-dimensional phase space. This combined noise space is bounded by \( p_e \in \{p_e^{\min}, p_e^{\max}\} = \{5 \cdot 10^{-4}, 3 \cdot 10^{-3}\} \) and \( p_l \in \{p_l^{\min}, p_l^{\max}\} = \{10^{-2}, 10^{-1}\} \), as used to produce Figures~\ref{fig:logical error rates} and~\ref{fig:all_angles_ler}. We sweep across it along lines of constant angle $\theta$ where $\theta$ is taken from a set of equally distributed angles from the origin of the phase space parameterised by polar coordinates. Each sweep line is linearly parameterised by a variable $x$, the \textit{noise scaling factor}, such that the corresponding mapping between fusion noise and $x$ is given by

\begin{equation}
\begin{split}
    p_e = (x\tilde{p_e} +p_e^{min})\cos{\theta}  \\
    p_l = (x\tilde{p_l} +p_l^{min})\sin{\theta} ,
\end{split}
\end{equation}

where $\tilde{p_e}=p_e^{max}-p_e^{min}$ and $\tilde{p_l}=p_l^{max}-p_l^{min}$.
Here $\theta=0,\pi/2$ are lines of error and erasure only.
With an i.i.d. (independent and identically distributed) model of both error and erasure of the fusions, we run $10^5$ Monte Carlo simulations with the modified union-find decoder (see section~\ref{suppl_decode} for details). Data from the complete simulation result for all noise combination sweeps is shown in Figure~\ref{fig:all_angles_ler}, where the error bars shown represent $\pm$ 1 standard deviation about the mean considering Poissonian noise due to the finite number of samples.

Erasure occurs with probability $p_l$ and corresponds to a measurement with unknown outcome. Thus, we can assume that the corresponding qubit has an error with probability $0.5$ \cite{Delfosse2020}. When there is no erasure (probability $1-p_l$), an error occurs with the Pauli error rate $p_e$. The overall error rate is therefore:

\begin{equation}
    p= \frac{p_l}{2}+(1-p_l)p_e.
\end{equation}

We define the concept of a pseudo-threshold as the physical noise rate $p$ when the probability that at least one of the $k$ logical qubits has an error $\bar{p}$ is equal to the probability that one of the $k$ physical qubits has an error $p_k^{phys}=1-(1-p)^k$. We refer to this as the \textit{break-even equation} and it is plotted as grey curves in Figure~\ref{fig:logical error rates} and Figure~\ref{fig:all_angles_ler}.

To produce Figure~\ref{fig:logical error rates}(c), we sweep an evenly distributed grid of points in the noise phase-space and similarly run the decoder on $10^5$ samples of each point. From this simulation, we construct a two-dimensional color plot showing $\bar{p}$ as a function of $p_e, p_l$ where the Voronoi cell around every grid point is assigned its value of $\bar{p}$.

\subsection{Decoder details}\label{suppl_decode}
To determine threshold error and loss rates of the foilated qLDPC codes, we need a decoder that can handle losses/erasures and also works for non-topological qLDPC codes. We chose a modified version of the union-find decoder \cite{Delfosse2021}, which was presented in Ref.~\cite{, uf-decoder}.

The original union-find decoder~\cite{Delfosse2021} works in two steps: first, it grows clusters on the Tanner graph starting from erasures and non-zero syndromes until all clusters are valid, where valid means that a cluster is decodable by only qubits within the cluster. For topological codes this so-called \textit{syndrome validation} is easy as a cluster is valid/decodable if and only if it has an even parity of non-zero syndromes~\cite{Delfosse2021}. Secondly, clusters are decoded, which for topological codes, works with the peeling-forest decoder from Ref.~\cite{Delfosse2020}.

The studied qLDPC codes~\cite{bicyclecodes} are non-topological and, therefore, neither syndrome validation by checking the cluster parity nor the peeling forest decoder can be applied. For this reason, we employ a modified union-find decoder~\cite{uf-decoder} that performs syndrome validation and decoding by solving the equation $\sigma=H_r\cdot e$ via Gaussian elimination. Here $\sigma$ is an $\mathbb{F}_2$ vector of the cluster's syndromes, $e$ is an $\mathbb{F}_2$ vector with the cluster's data qubits, and $H_r$ is the parity check matrix corresponding exclusively to this cluster. When a solution is found, it directly provides a decoding for the cluster. If there is no solution, the cluster cannot be decoded and is grown further. The cluster growth is performed node-by-node following breadth-first Tanner graph traversal, starting from erased qubits and non-zero syndromes. Cluster growth starting from erased qubits is performed first. Afterwards, we sort the nodes listed for subsequent cluster growth such that those belonging to invalid clusters are listed first. This sorting is the only difference to Algorithm 3 from Ref.~\cite{uf-decoder} and improves the logical error rates in the regime of high erasure rates. Finally, the subsequent cluster growth is performed in double steps on the Tanner graph such that cluster boundaries are syndrome checks and not data qubits~\cite{uf-decoder}. This makes sure that decoding a cluster does not flip syndromes that are not part of it.

\section*{Data availability}

The data generated for this article is openly available at~\footnote{Github repository https://github.com/susanxschen/qldpc-fusion-lattices.}.

\section*{Code availability}

The code written for this article is openly available at~\footnotemark[\value{footnote}].

\section*{Acknowledgements}

Part of this work was carried out using the computational facilities of the Advanced Computing Research Centre, University of Bristol - http://www.bristol.ac.uk/acrc/.We gratefully acknowledge financial support from Danmarks Grundforskningsfond (DNRF 139, Hy-Q Center for Hybrid Quantum Networks), and Danmarks Innovationsfond (IFD1003402609, FTQP). M.L.C acknowledges funding from Danmarks Innovationsfond (Grant No.~4298-00011B). S.X.C. acknowledges support from UK EPSRC (EP/SO23607/1).  S.P. acknowledges funding from the VILLUM FONDEN research grants No.~VIL50326 and No.~VIL60743, and support from the NNF Quantum Computing Programme. 

\section*{Author contributions}

S.P. and A.S.S. conceived the idea. S.P. supervised the project. S.X.C. performed the simulations and analysis. M.C.L. developed and incorporated the use of the decoder. M.L.C. contributed to the implementation of the repeat-until-success protocol. M.C.L. and S.X.C. further developed the repeat-until-success strategy. All authors discussed the results. S.X.C and M.C.L created the figures and wrote the manuscript with inputs from all authors. All authors have read and approved the manuscript.

\section*{Competing interests}

M.C.L and M.L.C are employees of Sparrow Quantum, a company that commercialises single-photon sources. The other authors do not have competing interests.

\makeatletter
\def\bibsection{\section*{References}}
\makeatother

\bibliography{biblio.bib}

\clearpage
\newpage

\makeatletter
\def\maketitle{
\@author@finish
\title@column\titleblock@produce
\suppressfloats[t]}
\makeatother

\title{Supplementary Information - Fusion-based implementation of qLDPC codes with quantum emitters}

\author{Susan X. Chen}
\email[Corresponding author: ]{susan.chen@nbi.ku.dk}
\affiliation{Quantum Engineering Centre for Doctoral Training, H. H. Wills Physics Laboratory and School of Electrical, Electronic, and Mechanical Engineering, University of Bristol, BS8 1FD, United Kingdom}
\affiliation{NNF Quantum Computing Programme, Niels Bohr Institute, University of Copenhagen, Blegdamsvej 17, DK-2100 Copenhagen Ø, Denmark}

\author{Matthias C. L\"{o}bl}
\affiliation{Center for Hybrid Quantum Networks (Hy-Q), The Niels Bohr Institute, University~of~Copenhagen,  DK-2100  Copenhagen~{\O}, Denmark}
\affiliation{Sparrow Quantum, Blegdamsvej 104A, DK-2100  Copenhagen~{\O}, Denmark}

\author{Ming Lai Chan}
\affiliation{Center for Hybrid Quantum Networks (Hy-Q), The Niels Bohr Institute, University~of~Copenhagen,  DK-2100  Copenhagen~{\O}, Denmark}
\affiliation{Sparrow Quantum, Blegdamsvej 104A, DK-2100  Copenhagen~{\O}, Denmark}

\author{Anders S. Sørensen}
\affiliation{Center for Hybrid Quantum Networks (Hy-Q), The Niels Bohr Institute, University~of~Copenhagen,  DK-2100  Copenhagen~{\O}, Denmark}

\author{Stefano Paesani}
\email{stefano.paesani@nbi.ku.dk}
\affiliation{NNF Quantum Computing Programme, Niels Bohr Institute, University of Copenhagen, Blegdamsvej 17, DK-2100 Copenhagen Ø, Denmark}

\maketitle

\onecolumngrid

\setcounter{equation}{0}
\setcounter{figure}{0}
\setcounter{table}{0}
\setcounter{page}{1}
\makeatletter
\renewcommand{\theequation}{S\arabic{equation}}
\renewcommand{\thefigure}{S\arabic{figure}}
\renewcommand{\thetable}{S\arabic{table}}
\renewcommand{\@seccntformat}[1]{%
  \csname the#1\endcsname.\quad
}

\setcounter{section}{0}
\setcounter{secnumdepth}{1}

\section{Details on the spin-based architecture with repeat-until-success encoded fusions} \label{supp:rus}

In practice, we consider the foliated lattice to be built layer by layer. To establish connectivity within layers, we use repeat-until-success (RUS) encoded fusions~\cite{Gliniasty2024, chan2024tailoring, Lobl2024}. RUS fusions are encoded versions of physical fusions, where a physical fusion has the following success ($s$), failure ($f$) and erasure ($l$) probabilities:

\begin{equation}
    P_s = P_f = \frac{1}{2}\eta^2 \hspace{0.3cm} \text{and} \hspace{0.3cm} P_l = 1-\eta^2, 
\end{equation}

Here $1-\eta$ is the photon loss probability given the total setup efficiency $\eta$. Due to the probabilistic nature of fusions, RUS is used to increase the success probability of a bond being created by allowing for multiple attempts in the event of physical failures.

We consider RUS fusions where a maximum of $N$ repetitions can be performed. An RUS fusion has three different outcomes: (1) RUS fusion success, in which case a bond is established between the two spins, (2) RUS fusion failure, in which case there is no bond between the spins, (3) RUS fusion loss, in which case both spins need to be measured in the $Z$-basis, removing all their bonds. As a function of the photon loss probability the RUS fusion outcomes have the following probabilities~\cite{Gliniasty2024}:

\begin{itemize}
    \item The probability of RUS success, i.e., obtaining both outcomes, which corresponds to events in which $0\leq i<N$ fusion failures occur before a successful fusion is given by: $P_N^{\bar{s}} = P_s \sum_{i = 0}^{N-1}(P_f)^{i}$. In this case a connection between the two spins is established.
    \item The probability of RUS erasure, i.e., losing both outcomes, referring to events where the first $i$ fusions failed and the subsequent fusions have photon loss is: $P_N^{\bar{l}}=P_l \sum_{i = 0}^{N-1}(P_f)^{i}$. Here both spins need to be measured in the $Z$-basis.
    \item The probability of RUS failure, i.e., only obtaining one of the two outcomes, where all $N$ repeats result in fusion failures is: $P_N^{\bar{f}}=(P_f)^N$. In this case, the connection between the two spins is not established.

\end{itemize}

We consider strategies for RUS sampling and scheduling, these can be understood with an approach from Ref.~\cite{PhysRevA.97.030301} to handle absent bonds in cluster state lattices.

\begin{figure}[h]
    \centering
    \includegraphics[width=0.95\textwidth]{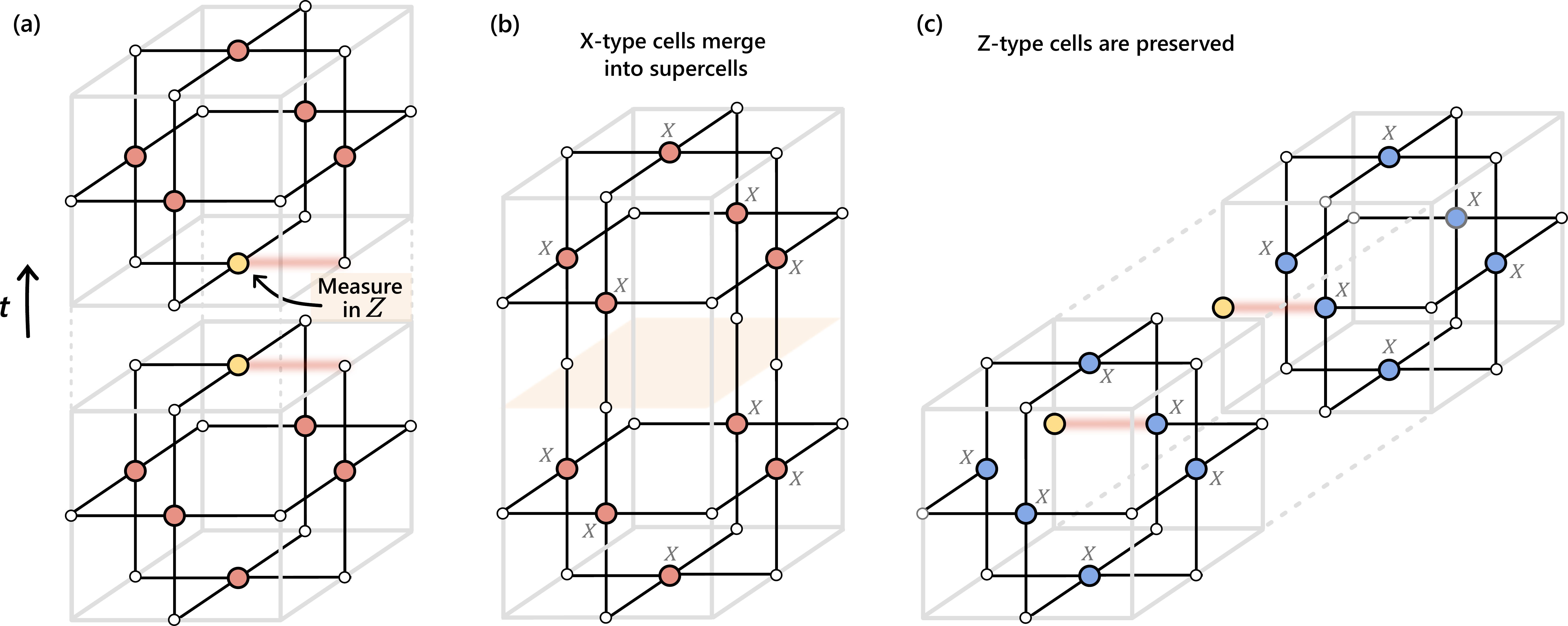}
    \caption{Illustration of the missing bond strategy for topological cluster state lattices in Ref.~\cite{PhysRevA.97.030301} (a) Two neighbouring X-type detector cells of $X^{\otimes6}$ on face qubits. If a bond is missing (marked in red), in this case we may choose to measure the (yellow) $X$-ancilla qubit in the $Z$-basis. The effect on the $X$-type lattice (b) is a supercell formed by merging the two cells, while on the $Z$-type lattice (c) is nothing and the cells stay intact.}
    \label{fig:missing-bond}
\end{figure}

Consider, for instance, the $X$-type lattice (the same reasoning applies to the $Z$-type lattice) - if a bond is missing, the two detector cells dependent on that bond (in both of the lattices) could be multiplied together to form a higher-weight \textit{supercell}~\cite{PhysRevLett.105.200502}, thereby removing the cells' dependencies on the missing bond. One may, however, measure the qubit at either end of the missing bond in the $Z$-basis to prevent the merging of cells in one of the two lattices. We discard the measurement outcome and therefore this effectively is the same action as erasing the said qubit \cite{PhysRevA.97.030301}. Figure ~\ref{fig:missing-bond}(a) shows the presence of a missing bond (indicated in red) with respect to $X$-type detector cells, the choice to measure the $X$-ancilla qubit (shown in yellow) merges the X-type cells as expected (Fig.~\ref{fig:missing-bond}(b)) while keeping the $Z$-type cells un-merged or  \textit{preserved} (Fig.~\ref{fig:missing-bond}(c)). Conversely, measuring the data qubit on the other end of the bond preserves the $X$-type cells and merges the $Z$-type cells.

We now consider the two strategies to handle imperfect RUS fusion. The first one, we refer to as \textit{standard} RUS (and described in Ref~\cite{Gliniasty2024}) and the second is the modified variant we devise and use in simulations. 

 \begin{enumerate}
     \item All fusions within one layer are executed simultaneously. Upon fusion erasure, which dephases both spins adjacent to the fusion photons, we measure both spins in the $Z$-basis, which effectively erases them. This causes one merge in both the $X$ and $Z$-type lattices. If we have a missing bond due to fusion failure, a $Z$-basis measurement is exclusively performed on the ancilla qubit. This strategy is equivalent to the one used in Ref~\cite{Gliniasty2024}.
     \item All fusions are executed sequentially, where previous measurements inform subsequent fusions. A fusion may be omitted if either of the spins involved already needs to be measured in the $Z$-basis due to a previous fusion erasure or failure. Omitting these unnecessary fusions improves the loss tolerance. Our treatment of fusion erasure is the same as above, however for RUS fusion failure, we instead randomly choose the spin at either end of the missing bonds to be measured in $Z$.
 \end{enumerate}

\begin{figure}[ht]
    \centering
\includegraphics[width=0.8\textwidth]{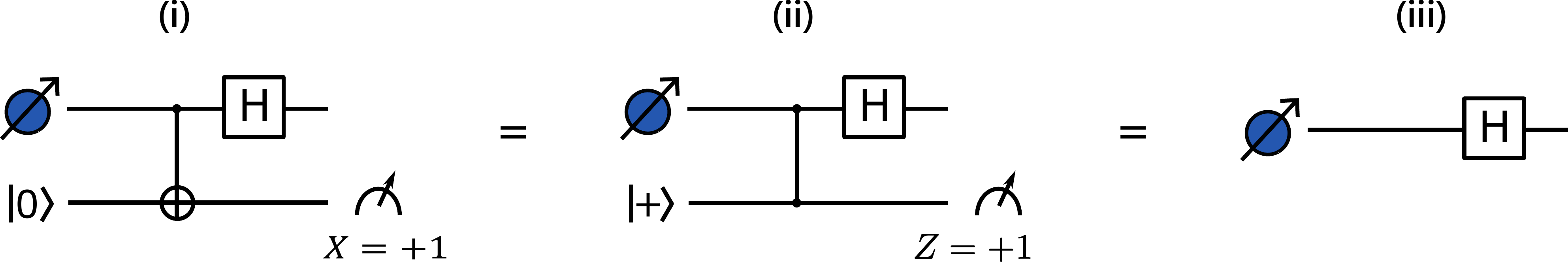}
    \caption{Equivalent circuits to move spins (blue) from one layer of the lattice to the next. While ordinarily (i) a photon would be emitted and measured in $X$ in the foliation picture, this is equivalent to (ii) a graph state with a dangling bond being measured in $Z$ which is the same as (iii) omitting its generation.}
    \label{fig:virtual-qubit}
\end{figure}

We note that after all bonds in a given layer have been attempted by RUS fusions, $H$ gates are applied to the spins to progress to the next layer. This can be understood by simultaneously executing the lattice construction and the teleportation of quantum information to the next layer by $X$-measurements. In Figure 1, a photon is first generated in place of the spin as a chain node. To teleport the encoded state forward in the lattice, this photon will subsequently be measured in the $X$-basis~\cite{Raussendorf2003}. However, this circuit is equivalent to a circuit in which the photon is never generated (Fig.~\ref{fig:virtual-qubit}). This equivalence is important, as we therefore can omit the photon generation. These virtual qubits are thus never generated and so the only photons that can suffer loss are those involved in fusions.

\begin{figure}[ht]
    \centering
    \includegraphics[width=0.7\linewidth]{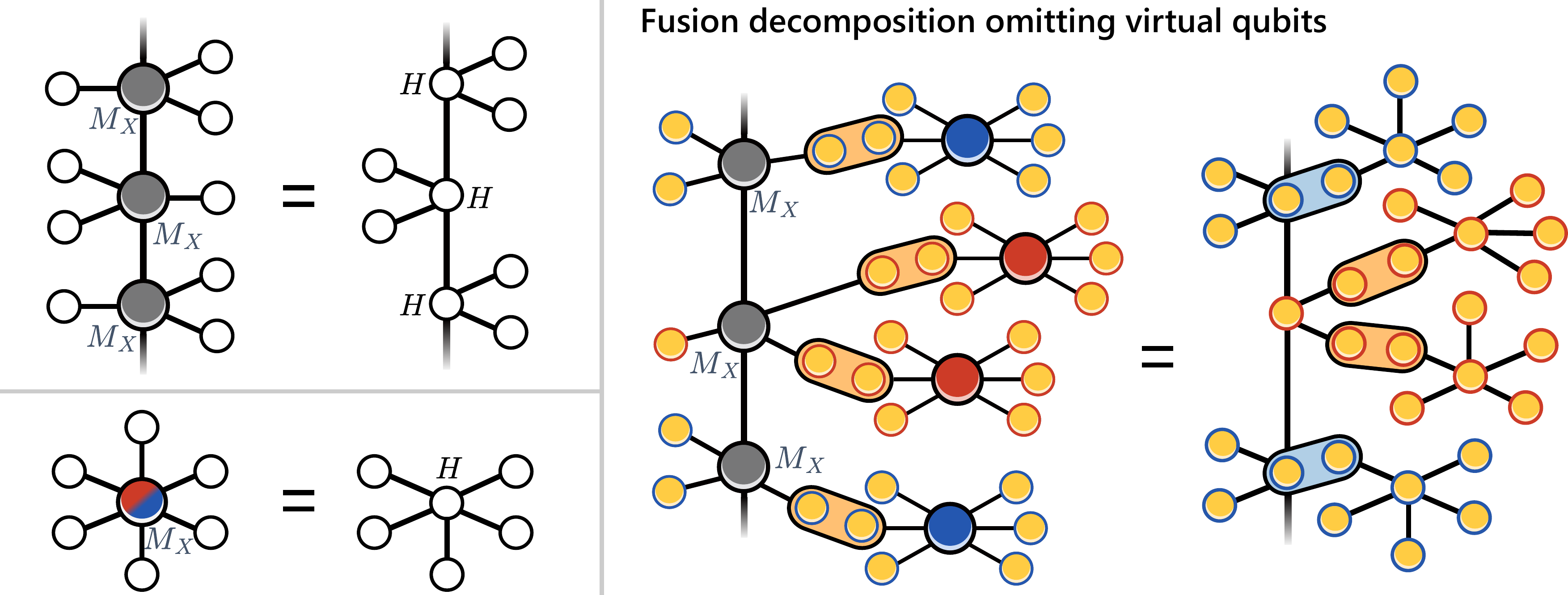}
    \caption{Equivalent resource states having omitted virtual qubits via $X$ measurement projection ($M_X$), here $H$ indicates a Hadamard operation on the qubit. Fusion operations shown may also incur basis changes as a result, for instance, the blue ovals indicate the rotation of the original basis by a Hadamard gate.}
    \label{fig:lattice without virtual qubits}
\end{figure}

In Figure~\ref{fig:lattice without virtual qubits}, we show the reduced resource states after the $X$-basis projection of these qubits. The branched chains reduce down to branched chains of one less leaf per node, $(\eta_Q/2 -1)$ leaves instead of $\eta_Q /2$, and the $(\eta_A + 1)$-star graphs reduce down to $\eta_A$-star graphs, while acquiring single qubit $H$ gates. The resulting lattice with virtual qubits omitted is shown in Figure~\ref{fig:lattice without virtual qubits} where some fusion bases (blue) are rotated by the $H$ gates.

\begin{figure}[ht]
    \centering
    \includegraphics[width=1.01\textwidth]{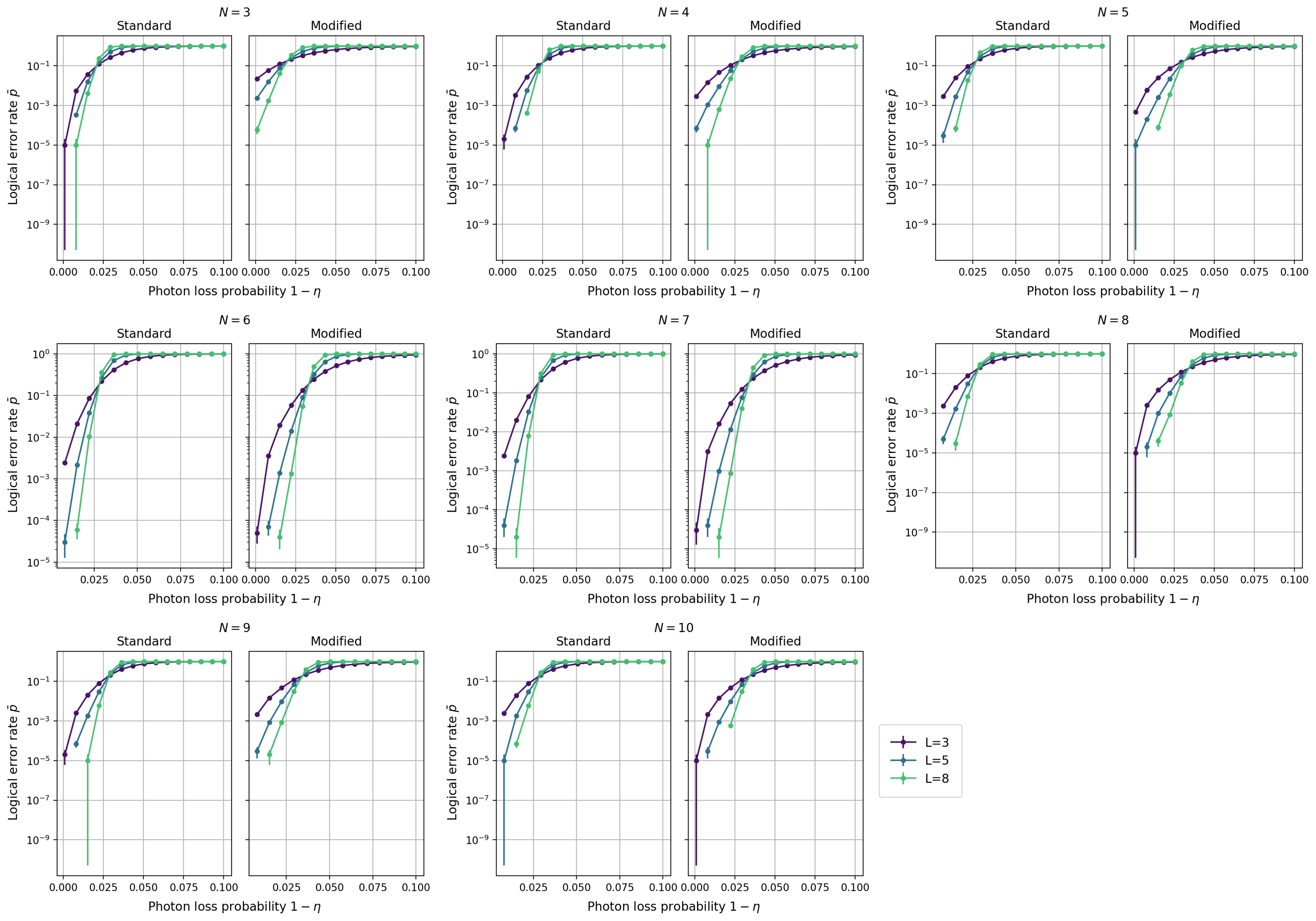}
    \caption{{Logical error rates $\bar{p}$ against photon loss probability $1-\eta$ for fusion lattices constructed from Toric codes of sizes $\{3,5,8\}$ under repeat-until-success with $N$, the number of maximum repeats, increasing from 3 to 10. In particular, a comparison is drawn between the standard strategy described in Ref.~\cite{Gliniasty2024} and our updated modified strategy inspired by Ref.~\cite{PhysRevA.97.030301}.}}
    \label{fig:rus-toric-data}
\end{figure}

\begin{figure}[ht]
    \centering
    \includegraphics[width=\textwidth]{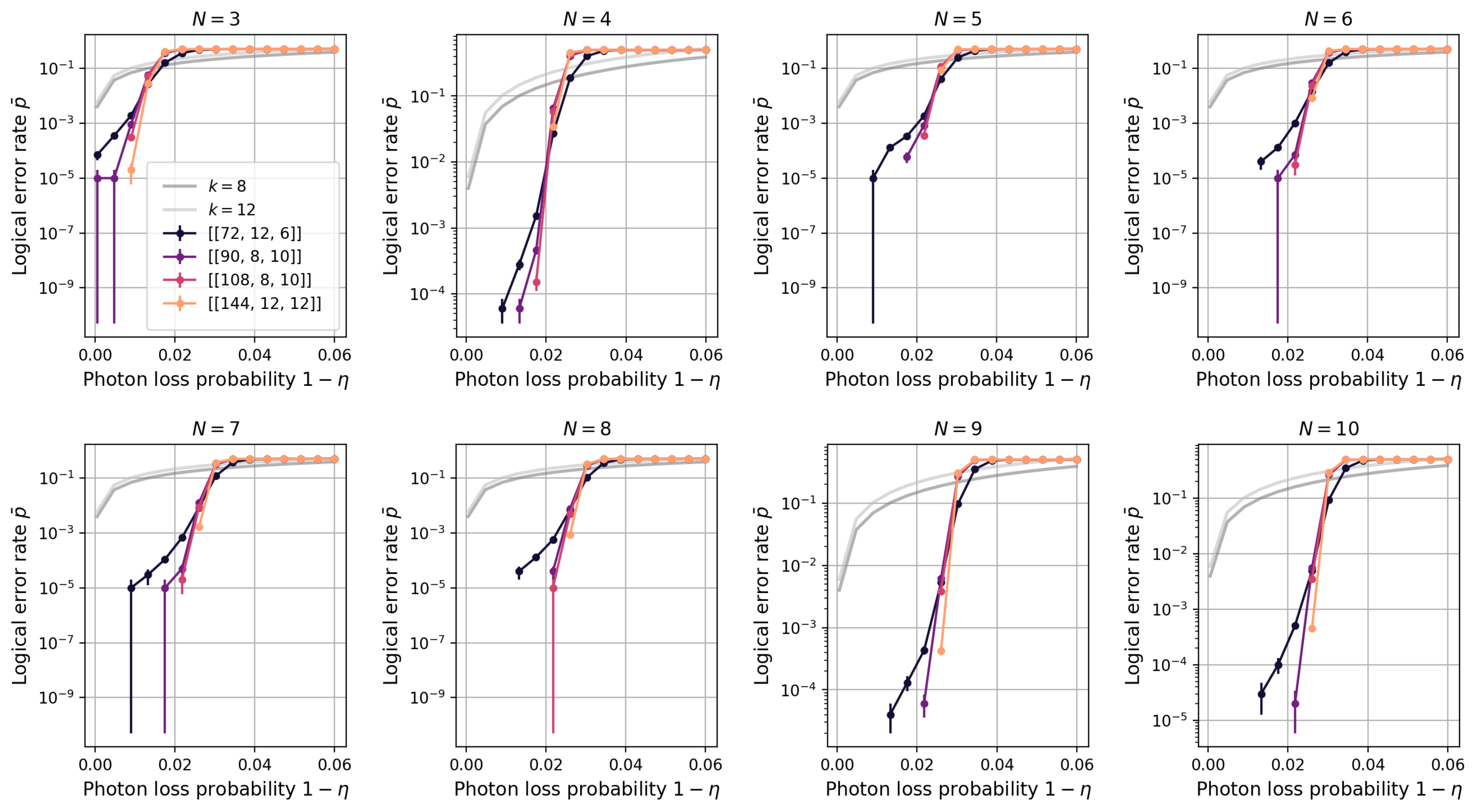}
    \caption{{Logical error rates $\bar{p}$  against photon loss probability $1-\eta$ for fusion lattices constructed from small examples of Bivariate Bicycle qLDPC codes under repeat-until-success with $N$, the number of maximum repeats, increasing from 3 to 10. As previously, the grey curves are the break-even equations for $k=8,12$.}}
    \label{fig:rus-BB-data}
\end{figure}

The pseudo-thresholds and thresholds shown in Figure 3(b) are determined in the following way. Applying the modified RUS strategy, we numerically sample which encoded fusion outcomes are lost. Then we perform erasure decoding on the $X$-type lattice ($10^5$ Monte Carlo samples). Gaussian elimination is used to identify whether any of the $k$ logical qubits are lost by checking if they have support on any of the missing fusion outcomes. We show the full data ($N=3$ to $10$) obtained for Toric code and Bivariate Bicycle code lattices in Figures~\ref{fig:rus-toric-data} and ~\ref{fig:rus-BB-data}, respectively. Figure~\ref{fig:rus-toric-data} also presents a comparison between the standard method and our modified strategy, highlighting the visible improvement from the latter.

\section{Modified fusions in realistic spin-based architectures} \label{supp:rotatedfusions}

In our architecture, we consider the situation that the fusion photons are two dangling bonds (leave nodes) attached to two spins. For the fusion-based construction of the foliated lattice shown in Fig.~\ref{fig:diagrams} of the main text we need a type of fusion that, up to local gates does the following: it makes (1) a CZ gate between the two spins upon success (add a graph bond), and (2) does two $Z$-basis measurements on the fusion photons upon failure (i.e. removes the nodes corresponding to the fusion photons). The first condition needs to be fulfilled to obtain the desired connectivity, and the second one ensures that the fusion can be repeated without any undesired effect on the graph state.

\begin{figure*}
    \centering
\includegraphics[width=\textwidth]{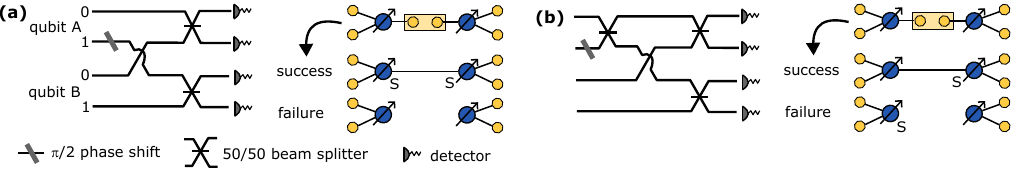}
    \caption{Two fusion setups for dual-rail encoded fusions. Up to local $S$-gates, both fusions have the desirable property to realize a connection between the two spins upon fusions success, while only removing the fusion photons upon fusion failure.}
    \label{fig:fusion}
\end{figure*}

Labelling the two fusion photons as $A,B$, the standard fusion measures the parities $X_AX_B, Z_AZ_B$ upon success and does the single-qubit measurements $Z_A, Z_B$ upon failure \cite{fbqc}. In this case, fusion failure corresponds to removing the fusion qubits as desired, but the fusion success case does not result in the desired spin-spin connectivity. This issue is illustrated in Fig. A5(b) in Ref.~\cite{Lobl2023c}. In Fig.~\ref{fig:fusion}(a,b) we give two exemplary setups for dual-rail-encoded fusion qubits that work for our purposes. The first one in Fig.~\ref{fig:fusion}(a) is the standard fusion setup with an additional phase-shifter implementing an $S$-gate ($S=\ket{0}\bra{0}+i\ket{1}\bra{1}$) on the first qubit. Upon success, this setup measures the parities $X_AY_B, Y_AX_B$. It can easily be seen that this measurement has the desired effect up to measurement-dependent Pauli gates (not mentioned in the following) and two $S$-gates on both spins (see section D.1 in Ref.~\cite{Lobl2023c}). Upon fusion failure, the fusion measures $Z_B, Z_B$ like the standard fusion and thus has the desired effect.

The second fusion setup is the standard fusion with an additional beam splitter and a phase-shifter implementing a Hadamard and an $S$-gate on the first qubit. This fusion measures the parities $X_AY_B, Y_AZ_B$ upon success. Up to an $S$-gate on the spin connected to qubit $B$, this gives the desired CZ-gate between the spins (see Fig. A5(d) in Ref.~\cite{Lobl2023c} or Appendix D.2 in Ref.~\cite{Lobl2024b}). Upon fusion failure, $Y_A, Z_B$ is measured which corresponds to removing the nodes $A, B$ and applying an $S$-gate to the spin connected to qubit $A$ (see e.g. Ref.~\cite{Hein2006}).

We finally remark that the additional $S$-gates commute with the gates applied to the spin as long as one only generates new photons entangled with the spin as dangling bonds (leaf nodes). This can be seen by considering the corresponding circuit from Fig.~\ref{fig:diagrams}(e) in the main text. This result is important from a practical point of view as it implies that the $S$-gates do not have to be actively removed between repeated fusions and not before an entire layer is finished.

\section{Bivariate Bicycle qLDPC codes}\label{supp bb codes}
CSS codes are defined by binary matrices $H_X$ and $H_Z$ of size $(n - k) \times n$, which specify complete sets of $X$-type and $Z$-type stabiliser generators, respectively, with each row indicating the qubits on which a distinct stabiliser acts. The two types of stabilisers must be mutually commuting, in other words, satisfy the orthogonality relation: 

\begin{equation}
    H_X H_Z^T=0 \hspace{0.2cm}\text{(mod 2)}
\end{equation}

Bivariate Bicycle qLDPC codes~\cite{bicyclecodes} are defined via the following parity check matrices 

\begin{equation}
    H_X = [A|B] \hspace{0.2cm} \text{ and } \hspace{0.2cm}  H_Z = [B^T|A^T] ,
\end{equation}
where $A$ and $B$ are sums of matrices $A=\sum_i^3 A_i$ and $B=\sum_i^3 B_i$, with $A_i$ and $B_i$ being powers of
 \begin{equation}
     x = S_l \otimes I_m \hspace{0.2cm} \text{ and } \hspace{0.2cm} y = I_l \otimes S_m ,
 \end{equation}
with $S_l$ being the cyclic shift matrix of size $l\times l$ and $I_m$ being the $m \times m$ identity matrix. A code defined as such will have a degree-6 Tanner graph, in contrast to 4 for the surface code. We select a number of small Bivariate Bicycle qLDPC code examples to construct and test our lattice constructions. These are shown in table \ref{tab:code examples}. One can prove, due to certain group properties, that these codes can be presented in a \textit{Toric layout}, meaning all vertices of the Tanner graph can be placed on a planar grid with periodic boundary conditions, like the Toric code~\cite{bicyclecodes}. In this layout, each vertex has horizontal and vertical edges connecting to its nearest neighbour vertices, forming a subset local embedding of the code, while the remaining two edges at each vertex correspond to non-local connections. A section of the Tanner graph of the smallest considered code $[[72, 12, 6]]$ is presented in its Toric layout in Figure 1(b) of the main text.

\begin{table}[h!]
    \centering
    \renewcommand{\arraystretch}{1.5} 
    \setlength{\tabcolsep}{9pt}
     \begin{tabular}{|c|c|c|c|c|c|}\hline
    \textbf{$[n, k, d]$} & \textbf{$l, m$} & \textbf{$A$} & \textbf{$B$} & \text{Error pseudo-threshold $(\%)$} & \text{Erasure pseudo-threshold $(\%)$} \\ \specialrule{1.5pt}{0pt}{0pt} 
    $[[72, 12, 6]]$ & $6, 6$ & $x^3 + y + y^2$ & $y^3 + x + x^2$ & 0.147 & 8.07 \\ \hline
    $[[90, 8, 10]]$ & $15, 3$ & $x^9 + y + y^2$ & $1 + x^2 + x^7$ & 0.158 & 8.41 \\ \hline
    $[[108, 8, 10]]$ & $9, 6$ & $x^3 + y + y^2$ & $y^3 + x + x^2$ & 0.176 & 8.53 \\ \hline
    $[[144, 12, 12]]$ & $12, 6$ & $x^3 + y + y^2$ & $y^3 + x + x^2$ & 0.181 & 8.70 \\ \hline
    \end{tabular}
    \caption{Examples of small Bivariate Bicycle qLDPC codes from Ref.~\cite{bicyclecodes} with their defining parameters and pseudo-thresholds of the corresponding fusion lattices.}
    \label{tab:code examples}
\end{table}

\end{document}